\documentclass[epj]{webofc}
\usepackage[utf8]{inputenc}
\usepackage[varg]{txfonts}   
\usepackage{booktabs}
\usepackage{xcolor}
\definecolor{darkred}{rgb}{0.4,0.0,0.0}
\definecolor{darkgreen}{rgb}{0.0,0.4,0.0}
\definecolor{darkblue}{rgb}{0.0,0.0,0.4}
\usepackage[bookmarks,linktocpage,colorlinks,
    linkcolor = darkred,
    urlcolor  = darkblue,
    citecolor = darkgreen]{hyperref}
\wocname{EPJ Web of Conferences}
\woctitle{Lattice2017}
\usepackage{braket}
\usepackage{units}
\usepackage{tikz}
\usetikzlibrary{fadings,shadows,positioning,calc,matrix,shapes,decorations.markings,shapes.arrows}

\graphicspath{{plots/}}

\DeclareMathOperator\tr{Tr}

\def\sl3c{\text{SL}(3,\mathbb{C})}
\def\su3{\text{SU(3)}}
\newcommand{\comment}[1]{}

\begin{document}
%
\selectlanguage{english}
\title{%
Selected inversion as key to a stable Langevin evolution across the QCD phase boundary
}
\author{%
\firstname{Jacques} \lastname{Bloch}\inst{1}\fnsep\thanks{Speaker, \email{jacques.bloch@ur.de}} \and
\firstname{Olaf} \lastname{Schenk}\inst{2}
}
\institute{%
University of Regensburg
\and
Università della Svizzera italiana, Lugano
}
\abstract{%
We present new results of full QCD at nonzero chemical potential. In PRD 92, 094516 (2015) the complex Langevin method was shown to break down when the inverse coupling decreases and enters the transition region from the deconfined to the confined phase. We found that the stochastic technique used to estimate the drift term can be very unstable for indefinite matrices. This may be avoided by using the full inverse of the Dirac operator, which is, however, too costly for four-dimensional lattices. The major breakthrough in this work was achieved by realizing that the inverse elements necessary for the drift term can be computed efficiently using the selected inversion technique provided by the parallel sparse direct solver package PARDISO. In our new study we show that no breakdown of the complex Langevin method is encountered and that simulations can be performed across the phase boundary.
}
\maketitle

\section{Introduction}

The lattice simulations of QCD at nonzero quark chemical potential are strongly hampered by the sign problem, caused by the complex fermion determinant. The complex Langevin (CL) method has drawn a lot of attention in recent years as a potential solution to this problem \cite{Aarts:2013bla}. Nevertheless, careful studies have shown that the method can break down or, even worse, can converge to the wrong solution, if the trajectories make excursions too far into the SL(3,$\mathbb{C}$) plane or come too close to a singularity of the drift \cite{Aarts:2011ax,Nagata:2016vkn}. Although conditions were derived that have to be satisfied for the CL solutions to be valid, the matching of these  conditions can only be verified a posteriori. Therefore, it cannot be excluded that the violation of the validity condition is due to numerical inaccuracies rather than to a theoretical deficiency of the method for the model being considered.

The first successful application of the CL method to QCD was made in the heavy dense approximation \cite{Sexty:2013ica}. For full QCD, the method was shown to work correctly in the deconfined phase, when the inverse coupling $\beta$ is large enough; however, it breaks down when $\beta$ gets smaller and the system crosses the phase boundary, such that no solutions are found in the confined phase \cite{Fodor:2015doa}. Other studies at larger $\beta$ and larger volumes also seem to converge to incorrect solutions \cite{Sinclair:2016nbg}. Recently there were suggestions to modify the CL evolutions through dynamical stabilization \cite{Jaeger2017} or deformation \cite{Shimasaki2017}, but the extrapolations needed to recover the original theory are not yet well controlled.

\begin{figure}
\centering
\begin{tikzpicture}
\node (phasediag) at (0,0) {\includegraphics[scale=0.35]{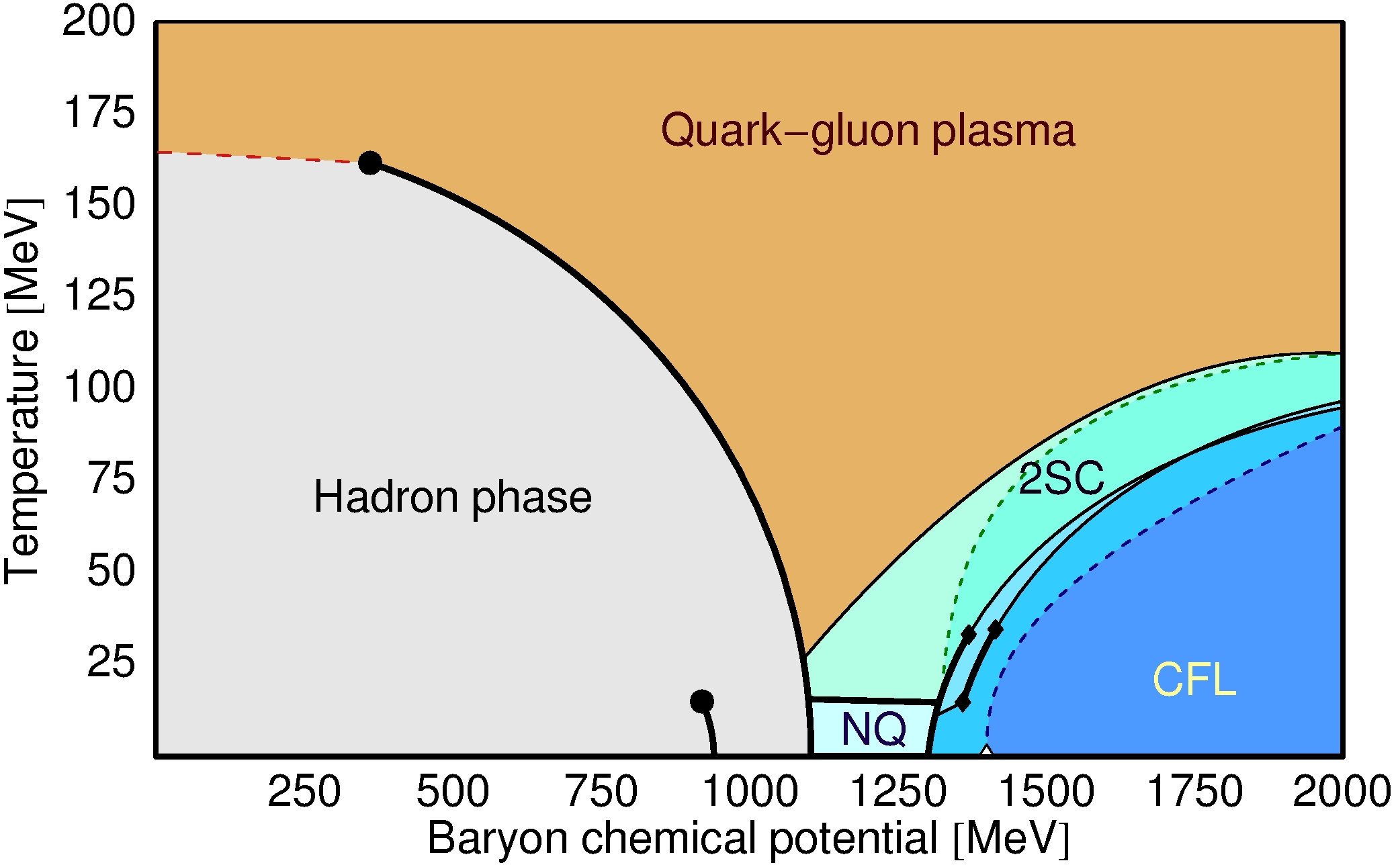}};
\node (pd1) [below right=8.5mm and 18mm] at (phasediag.north west) {};
\node[single arrow,draw=black,fill=black!30,minimum height=1.6cm,rotate=245] at (pd1) {};
\node[below left=-0.7mm and -5mm,color=red,thick] at (pd1) {\textbf{\large CL?}};
\draw[red,thick] ($(pd1)+(-0.1,-0.2)$) circle [radius=7mm];
\end{tikzpicture}
\caption{\label{phasediagram1}Sketch of the QCD phase diagram as a function of baryon chemical potential and temperature. The simulations of \cite{Fodor:2015doa} are performed at $\mu/T=1$ with decreasing $\beta$, which follows the gray arrow across the roof of the phase transition.\vspace{-3mm}}
\end{figure}
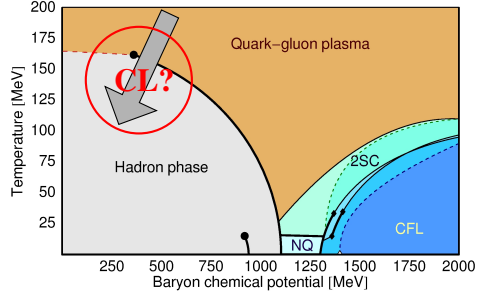

\begin{figure}
\centerline{
\hspace{-0.5mm}\includegraphics[width=0.35\textwidth]{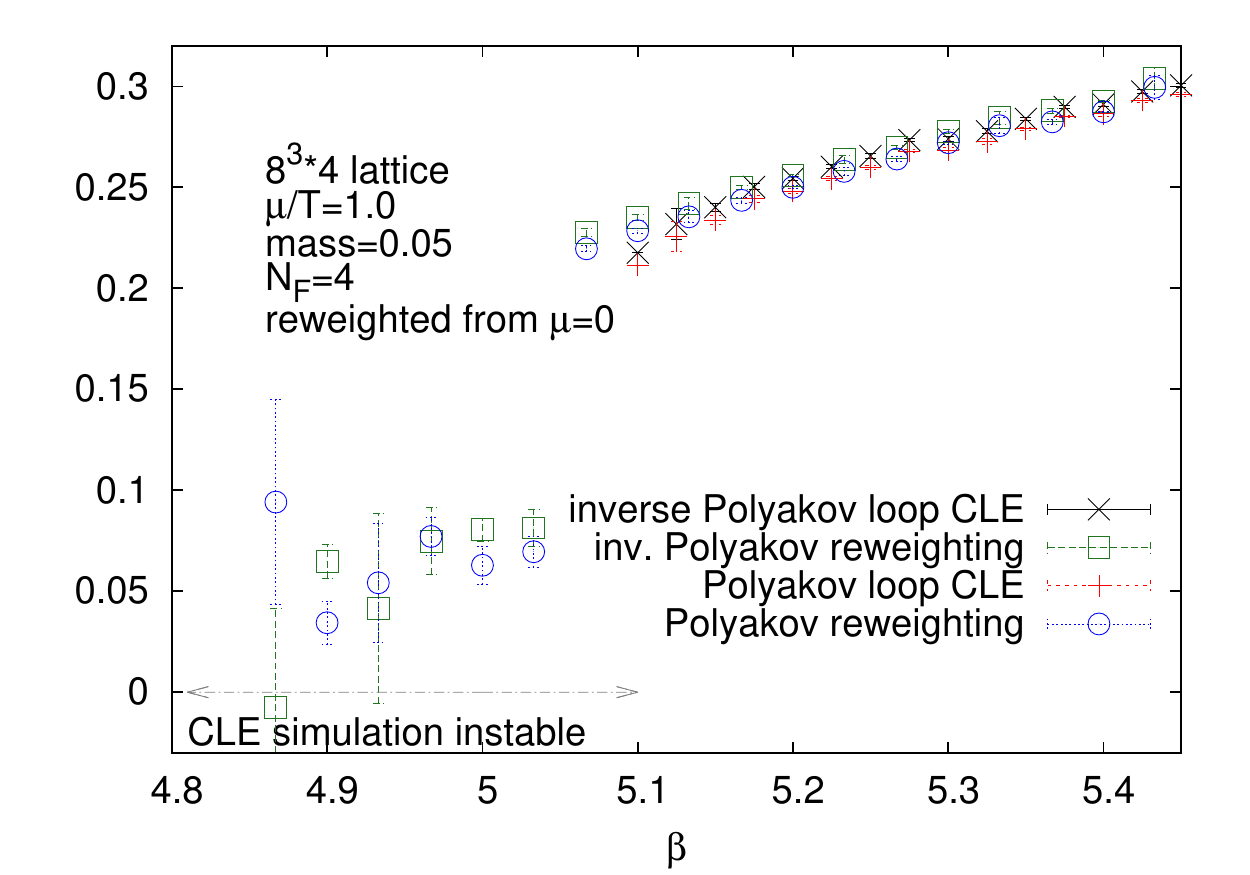}\hspace{-2mm}
\includegraphics[width=0.35\textwidth]{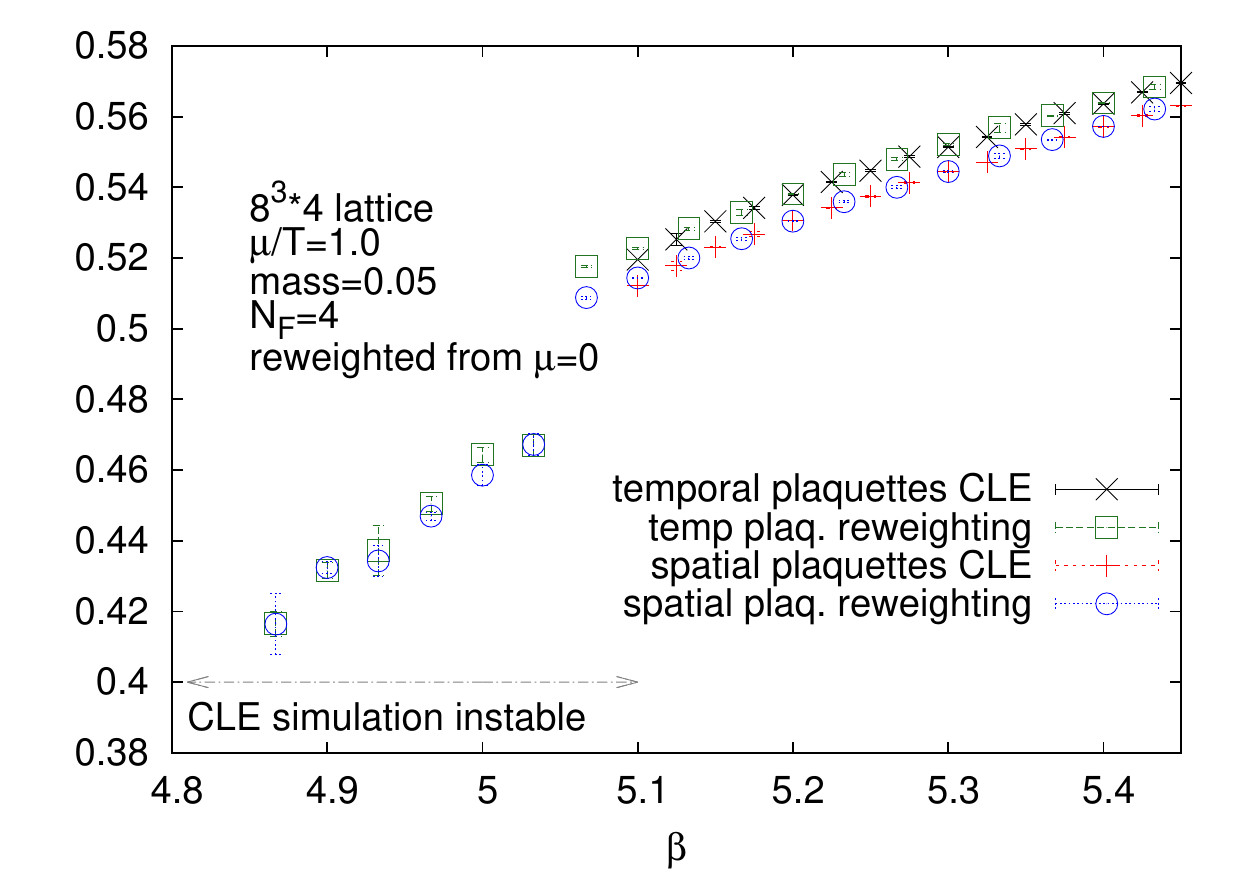}\hspace{-3mm}
\includegraphics[width=0.35\textwidth]{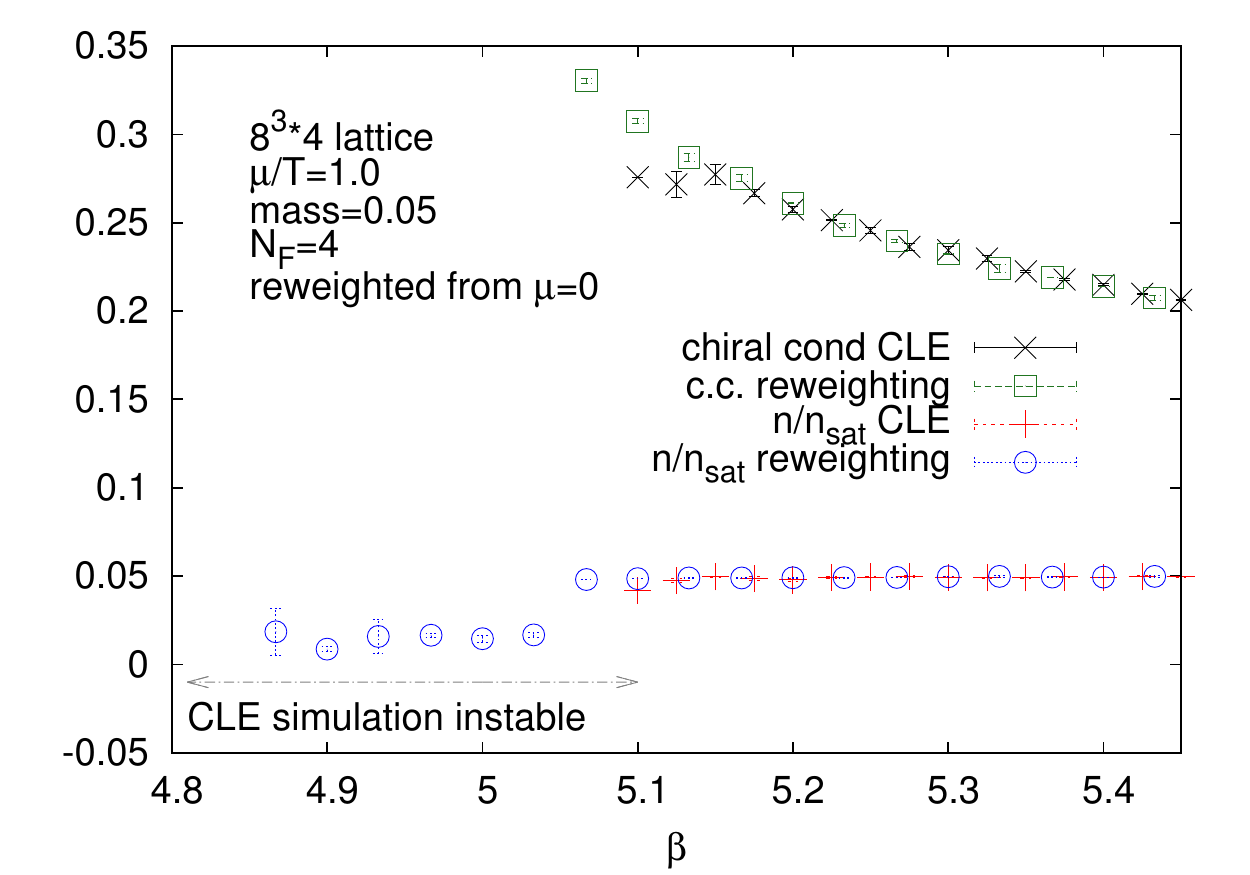}
}
\caption{\label{plotsPRD}Reproduction of Figs.~6 and 7 of \cite{Fodor:2015doa} showing results for the Polyakov loop and its inverse, the temporal and spatial plaquettes, and the chiral condensate and quark number density versus $\beta$ on an  $8^3\times 4$ lattice at $\mu/T=1$ and $m=0.05$. The CL results are compared with those obtained with reweighting from the $\mu=0$ ensemble. For $\beta<5.1$, no CL results are shown as the CL method became unstable.\vspace{-3mm}}
\end{figure}

Figure \ref{phasediagram1} gives a sketch of the QCD phase diagram as a function of temperature and chemical potential. The simulations reported by Fodor et al.~\cite{Fodor:2015doa} follow the gray arrow through the roof of the phase transition; however, the CL simulations break down when crossing the phase boundary, and no results were found inside the confined phase. In these simulations, the temperature was lowered by decreasing $\beta$ on an $8^3\times 4$ lattice with $\mu/T=1$ and $m=0.05$. For these parameter values the critical temperature corresponds to $\beta_c \approx 5.04$ at $\mu=0$. The results for the Polyakov loop and its inverse, the temporal and spatial plaquettes, the chiral condensate, and the quark number density published in \cite{Fodor:2015doa} are reproduced in Fig.\ \ref{plotsPRD}. In these plots the CL results are compared with data reweighted from the $\mu=0$ ensemble, however, CL results are only available for $\beta \geq 5.1$ as the CL simulations became unstable below this value. 

In this presentation we will show that the breakdown observed in \cite{Fodor:2015doa} can be cured and stable CL solutions can be found for smaller $\beta$ when the drift is computed exactly, rather than being estimated with stochastic techniques.

\section{Complex Langevin for QCD}

The lattice QCD partition function is given by
\begin{align}
Z=\left[\prod_{x=1}^V\prod_{\nu=1}^{d} \int\! d U_{x\nu}\right]\, \exp[-S_g] \det D(m;\mu)
\end{align}
with  Wilson gauge action $S_g$, staggered Dirac operator $D$, and links
\begin{align}
U_{x\nu} = \exp\Bigg[i\sum_{a=1}^8 z_{ax\nu} \lambda_a \Bigg] 
\end{align}
with Gell-Mann matrices $\lambda_a$ and link parameters $z_{ax\nu}$.
After discretization of the Langevin time, the CL evolution of the links in $\sl3c$ is described by
\begin{align}
U_{x\nu}(t+1) = R_{x\nu} (t) \: U_{x\nu}(t) ,
\end{align}
where, in the stochastic Euler discretization, 
\begin{align}
R_{x\nu} = \exp\left[i\sum_a \lambda_a (\epsilon K_{ax\nu} + \sqrt{\epsilon}\,\eta_{ax\nu})\right] \in \sl3c,
\end{align}
with Langevin step $\epsilon$ and Gaussian noise $\eta_{ax\nu}$.
The evolution is driven by the drift
\begin{align}
K_{ax\nu}  = - \partial_{ax\nu} S = K_{ax\nu}^\text{g} + K_{ax\nu}^\text{f}
\end{align}
with complex action $S=S_g - \log \det D$. In the following we will focus on the fermionic drift,
\begin{align}
K_{ax\nu}^\text{f} = \tr\left[D^{-1}\partial_{ax\nu} D\right],
\label{eq:fermdrift}
\end{align}
where $\partial_{ax\nu} D$ is the partial derivative of $D$ wrt the variables $z_{ax\nu}$.

\section{Fermionic drift term and selection inversion}

When looking at the breakdown of the CL method in \cite{Fodor:2015doa} it is not immediately clear if the violation of the CL validity conditions is genuine or rather due to numerical approximations. 

The traces in the fermionic drift \eqref{eq:fermdrift} involve the inverse Dirac operator. Computing the full inverse, for example, with Lapack, is too expensive for four-dimensional QCD, both in terms of CPU time and storage space. Until now, the explicit computation of $D^{-1}$ was avoided by using \emph{stochastic estimators} for the traces, 
\begin{align}
\tr [D^{-1}\partial_{ax\nu} D] \approx \eta^\dagger D^{-1} \partial_{ax\nu} D \eta .
\label{stochastic}
\end{align}
Although the merit of this technique is undisputed for positive-definite matrices, it can be problematic when applied to \emph{indefinite} matrices, as the number of noise vectors needed to get a good and stable estimate may be extremely large. As current CL algorithms typically estimate traces using a single noise vector, they rely on choosing $\epsilon$ tiny, such that consecutive CL steps are highly correlated and effectively provide an improved estimator to the trace after many Langevin steps. A potential danger is that this strategy may \emph{destabilize} the discrete time evolution beyond repair.

To clarify this situation we decided to investigate how using exact traces affects the stability of the CL evolution. As already mentioned, the exact traces require the inverse of the Dirac matrix, which is too costly to compute in full. Below we will analyze the drift term further and show that a relatively new numerical method, called \emph{selected inversion}, can be used to compute the drift exactly, while saving CPU time and storage space. This method also allows us to use a larger $\epsilon$, as the very small $\epsilon$ values used before were only needed to stabilize the drift when using the stochastic technique.

\tikzset{
    ncbar angle/.initial=90,
    ncbar/.style={
        to path=(\tikztostart)
        -- ($(\tikztostart)!#1!\pgfkeysvalueof{/tikz/ncbar angle}:(\tikztotarget)$)
        -- ($(\tikztotarget)!($(\tikztostart)!#1!\pgfkeysvalueof{/tikz/ncbar angle}:(\tikztotarget)$)!\pgfkeysvalueof{/tikz/ncbar angle}:(\tikztostart)$)
        -- (\tikztotarget)
    },
    ncbar/.default=0.5cm,
}
\tikzset{%
   register/.style={rectangle,rounded corners,draw},%
   single_entry/.style={rectangle, rounded corners, draw, thick},%
   single_entry_not_broadcasted/.style={rectangle, rounded corners, draw},%
   direction_line/.style={->, line width=0.25mm},%
   reduction_line/.style={->, thick, dashed},%
   connection_line/.style={thick},%
   my_node/.style={midway, draw, circle, thick, inner sep=3pt,
fill=white},%
	square left brace/.style={ncbar=0.5cm},
	square right brace/.style={ncbar=-0.5cm},
}

Consider the fermionic drift $K^\text{f}_{ax\nu}$ of Eq.~\eqref{eq:fermdrift}. The derivative matrix $\partial_{ax\nu} D $ is zero except for two $3\times3$ blocks at the positions of the link $U_{x\nu}$, so that the drift looks like
\begin{center}
\begin{tikzpicture}
\matrix [matrix of math nodes,left delimiter=(,right delimiter=)] (lhs)
{
               \phantom{1} & \phantom{1} & \phantom{1} & \phantom{1} &
\phantom{1} & \phantom{1} & \phantom{1} & \phantom{1} \\[-1mm]
               \phantom{1} & \phantom{1} & \phantom{1} & \phantom{1} &
\phantom{1} & \phantom{1} & \phantom{1} & \phantom{1} \\[-1mm]
               \phantom{1} & \phantom{1} & \phantom{1} & \phantom{1} &
\phantom{1} & \phantom{1} & \phantom{1} & \phantom{1} \\[-1mm]
               \phantom{1} & \phantom{1} & \phantom{1} & \phantom{1} &
\phantom{1} & \phantom{1} & \phantom{1} & \phantom{1} \\[-1mm]
               \phantom{1} & \phantom{1} & \phantom{1} & \phantom{1} &
\phantom{1} & \phantom{1} & \phantom{1} & \phantom{1} \\[-1mm]
               \phantom{1} & \phantom{1} & \phantom{1} & \phantom{1} &
\phantom{1} & \phantom{1} & \phantom{1} & \phantom{1} \\[-1mm]
               \phantom{1} & \phantom{1} & \phantom{1} & \phantom{1} &
\phantom{1} & \phantom{1} & \phantom{1} & \phantom{1} \\[-1mm]
               \phantom{1} & \phantom{1} & \phantom{1} & \phantom{1} &
\phantom{1} & \phantom{1} & \phantom{1} & \phantom{1} \\[-1mm]
           };
           \node [above=1.6cm] at (lhs.south) {dense};
           \node [below=1mm] at (lhs.south) {$D^{-1}$};
           \node (times) [right=.4cm] at (lhs.east) {$\times$};
           \matrix [matrix of math nodes,left delimiter=(,right delimiter=), right=.4cm of times.east] (op)
           {
               \phantom{1} & \phantom{1} & \phantom{1} & \phantom{1} &
\phantom{1} & \phantom{1} & \phantom{1} & \phantom{1} \\[-1mm]
               \phantom{1} & \phantom{1} & \phantom{1} & \phantom{1} &
\phantom{1} & \phantom{1} & \phantom{1} & \phantom{1} \\[-1mm]
               \phantom{1} & \phantom{1} & \phantom{1} & \phantom{1} &
\phantom{1} & \phantom{1} & \phantom{1} & \phantom{1} \\[-1mm]
               \phantom{1} & \phantom{1} & \phantom{1} & \phantom{1} &
\phantom{1} & \phantom{1} & \phantom{1} & \phantom{1} \\[-1mm]
               \phantom{1} & \phantom{1} & \phantom{1} & \phantom{1} &
\phantom{1} & \phantom{1} & \phantom{1} & \phantom{1} \\[-1mm]
               \phantom{1} & \phantom{1} & \phantom{1} & \phantom{1} &
\phantom{1} & \phantom{1} & \phantom{1} & \phantom{1} \\[-1mm]
               \phantom{1} & \phantom{1} & \phantom{1} & \phantom{1} &
\phantom{1} & \phantom{1} & \phantom{1} & \phantom{1} \\[-1mm]
               \phantom{1} & \phantom{1} & \phantom{1} & \phantom{1} &
\phantom{1} & \phantom{1} & \phantom{1} & \phantom{1} \\[-1mm]
           };
          \draw[single_entry,color=red]  (op-3-6.north west) rectangle (op-4-7.north west) node[midway] {$F$};
          \draw[single_entry,color=blue]  (op-6-3.north west) rectangle (op-7-4.north west) node[midway] {$B$};

           \node [below=1mm] at (op.south) {$\partial_{ax\nu} D$};
			\node (i) [above right=5mm and 9.3mm] at (op.north west) {$x$};
			\node (i1) [below=1mm] at (i) {};
			\node (i2) [below=5mm] at (i1) {};
           \draw[->] (i1)  -- (i2);
			\node (id)[above right=5mm and 19.0mm] at (op.north west) {$x+\hat\nu$};
			\node (id1) [below=1mm] at (id) {};
			\node (id2) [below=5mm] at (id1) {};
           \draw[->] (id1)  -- (id2);

			\node (ih2) [below right=9mm and 5mm] at (op.north east) {};
			\node (ih1) [right=5mm] at (ih2) {};
			\node (ih) [right=-1mm] at (ih1) {$x$};
           \draw[->] (ih1)  -- (ih2);

			\node (idh2) [below right=20mm and 5mm] at (op.north east) {};
			\node (idh1) [right=5mm] at (idh2) {};
			\node (idh) [above right=-2mm and -1mm] at (idh1) {$x+\hat\nu$};
           \draw[->] (idh1)  -- (idh2);

	\node [left=24mm] at (lhs) {$K^\text{f}_{ax\nu}=\tr$};
	\draw [very thick] ($(lhs)+(-17mm,-19mm)$) to [square left brace] ($(lhs)+(-17mm,19mm)$);
	\draw [very thick] ($(op)+(17mm,-19mm)$) to [square right brace] ($(op)+(17mm,19mm)$);
\end{tikzpicture}
\end{center}
In the matrix product the sparse derivative matrix effectively selects out the two corresponding columns of $D^{-1}$ such that this can be rewritten as
\begin{center}
\begin{tikzpicture}
           \matrix [matrix of math nodes,left delimiter=(,right delimiter=)] (lhs)
           {
               \phantom{1} & \phantom{1} & \phantom{1} & \phantom{1} &
\phantom{1} & \phantom{1} & \phantom{1} & \phantom{1} \\[-1mm]
               \phantom{1} & \phantom{1} & \phantom{1} & \phantom{1} &
\phantom{1} & \phantom{1} & \phantom{1} & \phantom{1} \\[-1mm]
               \phantom{1} & \phantom{1} & \phantom{1} & \phantom{1} &
\phantom{1} & \phantom{1} & \phantom{1} & \phantom{1} \\[-1mm]
               \phantom{1} & \phantom{1} & \phantom{1} & \phantom{1} &
\phantom{1} & \phantom{1} & \phantom{1} & \phantom{1} \\[-1mm]
               \phantom{1} & \phantom{1} & \phantom{1} & \phantom{1} &
\phantom{1} & \phantom{1} & \phantom{1} & \phantom{1} \\[-1mm]
               \phantom{1} & \phantom{1} & \phantom{1} & \phantom{1} &
\phantom{1} & \phantom{1} & \phantom{1} & \phantom{1} \\[-1mm]
               \phantom{1} & \phantom{1} & \phantom{1} & \phantom{1} &
\phantom{1} & \phantom{1} & \phantom{1} & \phantom{1} \\[-1mm]
               \phantom{1} & \phantom{1} & \phantom{1} & \phantom{1} &
\phantom{1} & \phantom{1} & \phantom{1} & \phantom{1} \\[-1mm]
           };

          \draw[single_entry,color=red]  ($(lhs.north west)+(9mm,-2mm)$) rectangle  ($(lhs.north west)+(13mm,-30mm)$);

          \draw[single_entry,color=blue] ($(lhs.north west)+(22mm,-2mm)$) rectangle ($(lhs.north west)+(26mm,-30mm)$);

           \node [below=1mm] at (lhs.south) {$D^{-1}$};
           
			\node (i) [above right=5mm and 8.1mm] at (lhs.north west) {$C_x$};
			\node (i1) [below=1mm] at (i) {};
			\node (i2) [below=5mm] at (i1) {};
           \draw[->] (i1)  -- (i2);
			\node (id)[above right=5mm and 19.7mm] at (lhs.north west) {$C_{x+\hat\nu}$};
			\node (id1) [below=1mm] at (id) {};
			\node (id2) [below=5mm] at (id1) {};
           \draw[->] (id1)  -- (id2);

           \node (times) [right=.4cm] at (lhs.east) {$\times$};
           \matrix [matrix of math nodes,left delimiter=(,right delimiter=), right=.4cm of times.east] (op)
           {
               \phantom{1} & \phantom{1} & \phantom{1} & \phantom{1} &
\phantom{1} & \phantom{1} & \phantom{1} & \phantom{1} \\[-1mm]
               \phantom{1} & \phantom{1} & \phantom{1} & \phantom{1} &
\phantom{1} & \phantom{1} & \phantom{1} & \phantom{1} \\[-1mm]
               \phantom{1} & \phantom{1} & \phantom{1} & \phantom{1} &
\phantom{1} & \phantom{1} & \phantom{1} & \phantom{1} \\[-1mm]
               \phantom{1} & \phantom{1} & \phantom{1} & \phantom{1} &
\phantom{1} & \phantom{1} & \phantom{1} & \phantom{1} \\[-1mm]
               \phantom{1} & \phantom{1} & \phantom{1} & \phantom{1} &
\phantom{1} & \phantom{1} & \phantom{1} & \phantom{1} \\[-1mm]
               \phantom{1} & \phantom{1} & \phantom{1} & \phantom{1} &
\phantom{1} & \phantom{1} & \phantom{1} & \phantom{1} \\[-1mm]
               \phantom{1} & \phantom{1} & \phantom{1} & \phantom{1} &
\phantom{1} & \phantom{1} & \phantom{1} & \phantom{1} \\[-1mm]
               \phantom{1} & \phantom{1} & \phantom{1} & \phantom{1} &
\phantom{1} & \phantom{1} & \phantom{1} & \phantom{1} \\[-1mm]
           };

          \draw[single_entry,color=red]  (op-3-6.north west) rectangle (op-4-7.north west) node[midway] {$F$};
          \draw[single_entry,color=blue]  (op-6-3.north west) rectangle (op-7-4.north west) node[midway] {$B$};

           \node [below=1mm] at (op.south) {$\partial_{ax\nu} D$};
			\node (i) [above right=5mm and 9.3mm] at (op.north west) {$x$};
			\node (i1) [below=1mm] at (i) {};
			\node (i2) [below=5mm] at (i1) {};
           \draw[->] (i1)  -- (i2);
			\node (id)[above right=5mm and 19.0mm] at (op.north west) {$x+\hat\nu$};
			\node (id1) [below=1mm] at (id) {};
			\node (id2) [below=5mm] at (id1) {};
           \draw[->] (id1)  -- (id2);

			\node (ih2) [below right=9mm and 5mm] at (op.north east) {};
			\node (ih1) [right=5mm] at (ih2) {};
			\node (ih) [right=-1mm] at (ih1) {$x$};
           \draw[->] (ih1)  -- (ih2);

			\node (idh2) [below right=20mm and 5mm] at (op.north east) {};
			\node (idh1) [right=5mm] at (idh2) {};
			\node (idh) [above right=-2mm and -1mm] at (idh1) {$x+\hat\nu$};
           \draw[->] (idh1)  -- (idh2);

	\node [left=24mm] at (lhs) {$K^\text{f}_{ax\nu}=\tr$};
	\draw [very thick] ($(lhs)+(-17mm,-19mm)$) to [square left brace] ($(lhs)+(-17mm,19mm)$);
	\draw [very thick] ($(op)+(17mm,-19mm)$) to [square right brace] ($(op)+(17mm,19mm)$);

       \end{tikzpicture}
\end{center}
where $C_{x+\hat\nu} \times B$ and $C_x \times F$, respectively, give the columns $x$ and $x+\hat\nu$ of $D^{-1} \partial_{ax\nu} D$. Only two $3\times3$ blocks of $D^{-1}$ contribute to the trace of this matrix product, so this simplifies further to
\begin{center}
       \begin{tikzpicture}
       
           \matrix [matrix of math nodes,left delimiter=(,right delimiter=)] (lhs)
           {
               \phantom{1} & \phantom{1} & \phantom{1} & \phantom{1} &
\phantom{1} & \phantom{1} & \phantom{1} & \phantom{1} \\[-1mm]
               \phantom{1} & \phantom{1} & \phantom{1} & \phantom{1} &
\phantom{1} & \phantom{1} & \phantom{1} & \phantom{1} \\[-1mm]
               \phantom{1} & \phantom{1} & \phantom{1} & \phantom{1} &
\phantom{1} & \phantom{1} & \phantom{1} & \phantom{1} \\[-1mm]
               \phantom{1} & \phantom{1} & \phantom{1} & \phantom{1} &
\phantom{1} & \phantom{1} & \phantom{1} & \phantom{1} \\[-1mm]
               \phantom{1} & \phantom{1} & \phantom{1} & \phantom{1} &
\phantom{1} & \phantom{1} & \phantom{1} & \phantom{1} \\[-1mm]
               \phantom{1} & \phantom{1} & \phantom{1} & \phantom{1} &
\phantom{1} & \phantom{1} & \phantom{1} & \phantom{1} \\[-1mm]
               \phantom{1} & \phantom{1} & \phantom{1} & \phantom{1} &
\phantom{1} & \phantom{1} & \phantom{1} & \phantom{1} \\[-1mm]
               \phantom{1} & \phantom{1} & \phantom{1} & \phantom{1} &
\phantom{1} & \phantom{1} & \phantom{1} & \phantom{1} \\[-1mm]
           };

          \draw[single_entry,color=blue]  (lhs-3-6.north west) rectangle (lhs-4-7.north west) node[midway] {$P$};
          \draw[single_entry,color=red]  (lhs-6-3.north west) rectangle (lhs-7-4.north west) node[midway] {$Q$};

           \node [below=1mm] at (lhs.south) {$D^{-1}$};
           
			\node (i) [above right=5mm and 9.2mm] at (lhs.north west) {$x$};
			\node (i1) [below=1mm] at (i) {};
			\node (i2) [below=5mm] at (i1) {};
           \draw[->] (i1)  -- (i2);
			\node (id)[above right=5mm and 19.0mm] at (lhs.north west) {$x+\hat\nu$};
			\node (id1) [below=1mm] at (id) {};
			\node (id2) [below=5mm] at (id1) {};
           \draw[->] (id1)  -- (id2);

           \node (times) [right=.4cm] at (lhs.east) {$\times$};
           \matrix [matrix of math nodes,left delimiter=(,right delimiter=), right=.4cm of times.east] (op)
           {
               \phantom{1} & \phantom{1} & \phantom{1} & \phantom{1} &
\phantom{1} & \phantom{1} & \phantom{1} & \phantom{1} \\[-1mm]
               \phantom{1} & \phantom{1} & \phantom{1} & \phantom{1} &
\phantom{1} & \phantom{1} & \phantom{1} & \phantom{1} \\[-1mm]
               \phantom{1} & \phantom{1} & \phantom{1} & \phantom{1} &
\phantom{1} & \phantom{1} & \phantom{1} & \phantom{1} \\[-1mm]
               \phantom{1} & \phantom{1} & \phantom{1} & \phantom{1} &
\phantom{1} & \phantom{1} & \phantom{1} & \phantom{1} \\[-1mm]
               \phantom{1} & \phantom{1} & \phantom{1} & \phantom{1} &
\phantom{1} & \phantom{1} & \phantom{1} & \phantom{1} \\[-1mm]
               \phantom{1} & \phantom{1} & \phantom{1} & \phantom{1} &
\phantom{1} & \phantom{1} & \phantom{1} & \phantom{1} \\[-1mm]
               \phantom{1} & \phantom{1} & \phantom{1} & \phantom{1} &
\phantom{1} & \phantom{1} & \phantom{1} & \phantom{1} \\[-1mm]
               \phantom{1} & \phantom{1} & \phantom{1} & \phantom{1} &
\phantom{1} & \phantom{1} & \phantom{1} & \phantom{1} \\[-1mm]
           };

          \draw[single_entry,color=red]  (op-3-6.north west) rectangle (op-4-7.north west) node[midway] {$F$};
          \draw[single_entry,color=blue]  (op-6-3.north west) rectangle (op-7-4.north west) node[midway] {$B$};

           \node [below=1mm] at (op.south) {$\partial_{ax\nu} D$};
			\node (i) [above right=5mm and 9.2mm] at (op.north west) {$x$};
			\node (i1) [below=1mm] at (i) {};
			\node (i2) [below=5mm] at (i1) {};
           \draw[->] (i1)  -- (i2);
			\node (id)[above right=5mm and 19.0mm] at (op.north west) {$x+\hat\nu$};
			\node (id1) [below=1mm] at (id) {};
			\node (id2) [below=5mm] at (id1) {};
           \draw[->] (id1)  -- (id2);

			\node (ih2) [below right=9mm and 5mm] at (op.north east) {};
			\node (ih1) [right=5mm] at (ih2) {};
			\node (ih) [right=-1mm] at (ih1) {$x$};
           \draw[->] (ih1)  -- (ih2);

			\node (idh2) [below right=20mm and 5mm] at (op.north east) {};
			\node (idh1) [right=5mm] at (idh2) {};
			\node (idh) [above right=-2mm and -1mm] at (idh1) {$x+\hat\nu$};
           \draw[->] (idh1)  -- (idh2);

	\node [left=24mm] at (lhs) {$K^\text{f}_{ax\nu}=\tr$};
	\draw [very thick] ($(lhs)+(-17mm,-19mm)$) to [square left brace] ($(lhs)+(-17mm,19mm)$);
	\draw [very thick] ($(op)+(17mm,-19mm)$) to [square right brace] ($(op)+(17mm,19mm)$);
       \end{tikzpicture}
\end{center}
Each drift term can thus be written as
\begin{align}
K^\text{f}_{ax\nu} = \tr (P \cdot B) + \tr (Q\cdot F) 
\end{align}
and, hence, the computation of the drift \emph{only requires elements of $D^{-1}$ where $D$ itself is nonzero}. Note that, the same statement holds for the computation of the fermionic observables.

This observation is what brought us to consider the selected inversion technique to compute the drift term. The method consists of a sparse LU-factorization followed by a selected inversion, which \textit{exactly} computes \textit{selected} elements of the inverse $D^{-1}$ of a general matrix $D$. The subset of selected elements is defined by the set of \textit{nonzero entries} in $D$. The method is based on the fact that this specific subset of $D^{-1}$ can be evaluated without computing any inverse entry from outside of the subset. This is what substantially speeds up the computation compared to routines computing the full inverse. Moreover, as only the inverse elements of this subset are computed, they can be stored in sparse format. The parallel implementation of the selected inversion technique, as described in \cite{Kuzmin2013}, can be found in the latest version of the parallel sparse direct solver PARDISO \cite{Schenk2004}.

It is worthwhile to note that the selected inversion method is optimally used in the CL evolution, as the selected inversion method precisely yields all the elements of $D^{-1}$ needed for the drift terms and for the fermionic observables.

\begin{figure}
\centerline{\includegraphics{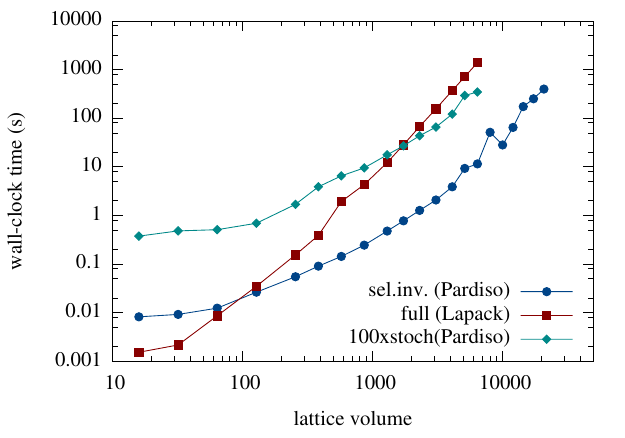}}
\caption{\label{fig:performance}Comparison of the wall-clock time versus lattice volume for one selected inversion of Pardiso, one full Lapack inversion, and 100 single-source inversions (also performed with the sparse Pardiso solver) for the stochastic trace estimator.\vspace{-5mm}}
\end{figure}

In contrast to the stochastic technique, the selected inversion allows us to compute the drift exactly, and this in a way that is much more efficient than Lapack, both in terms of CPU times and storage space. 
In Fig.\ \ref{fig:performance} we compare the scaling of the wall-clock time between the selected inversion from PARDISO, the full Lapack inversion, and the stochastic technique (where we used 100 intermediate steps due to the smaller choice of $\epsilon$ as explained after \eqref{stochastic}) as a function of the lattice volume. There is a clear performance gain when using the selected inversion, which is a factor of 100 compared to the Lapack dense inverse for a lattice size of $8^4$. The comparison with the stochastic technique is more delicate as the stability of the CL evolution can be affected by the latter and merely comparing timings does not tell the whole story.

The volume scaling of the selected inversion seems to be somewhat better than $N^3$. From other applications, the selected inverse is known to scale like $N^2$ for three-dimensional problems, however, as this is the first application to a four-dimensional problem the scaling has to be investigated further.

\section{Results}

\subsection{CL evolution for varying $\beta$}
\begin{figure}
\centerline{\includegraphics{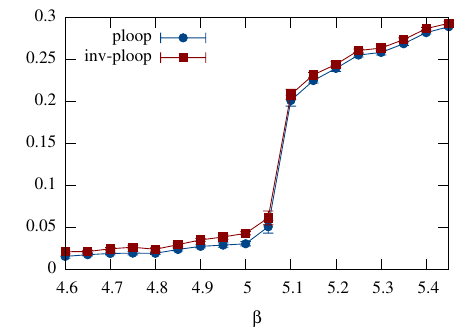}\includegraphics{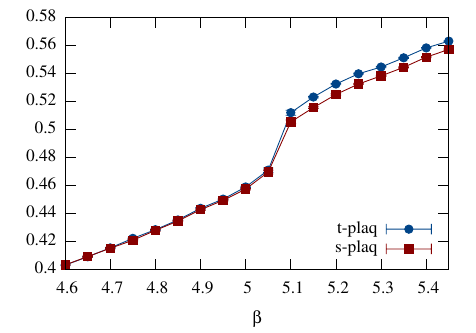}\includegraphics{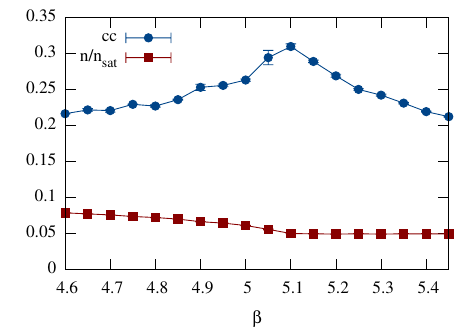}}
\caption{\label{fig:CLselinv1}Results of CL evolution with selected inversion for the Polyakov loop and inverse Polyakov loop (left), temporal and spatial plaquettes (middle), and chiral condensate and quark number density (right) as a function of $\beta$ for $m=0.05$ and $\mu=0.25$ on an $8^3\times4$ lattice.\vspace{-6mm}}
\end{figure}

In this section we present the first results of the CL method using the selected inversion.  The simulations were performed on the Xeon cluster at the ICS, Lugano. We performed the same study as in \cite{Fodor:2015doa} and investigate QCD across the phase boundary as sketched in Fig.~\ref{phasediagram1}, decreasing the temperature from the deconfined to the confined phase by varying $\beta$ for constant $\mu/T=1$ on an $8^3\times4$ lattice with $m=0.05$ ($\beta_c \approx 5.04$ at $\mu=0$). Whereas the CL method broke down below $\beta=5.1$ in the original study, we are now able to generate stable CL trajectories for all investigated $\beta$ values without any further tuning. Moreover, this was done using $\epsilon=0.001$, which is much larger than before.  The length of the trajectories is 30 Langevin time (Lt) of which 5 Lt are discarded as thermalization. From the results, shown in Fig.~\ref{fig:CLselinv1}, we see that the simulations perform well across the phase boundary, and the complete range from $\beta=5.45$ to $\beta=4.6$ can be simulated without any problem. The numerical implementation includes gauge cooling, to avoid excessive excursions in SL(3,$\mathbb{C}$), and an adaptive step size such that the continuum trajectory is properly followed, even when the drift is large. 

\begin{figure}[b]
\centerline{\includegraphics{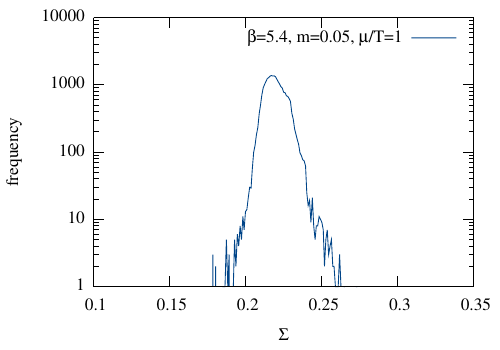}\includegraphics{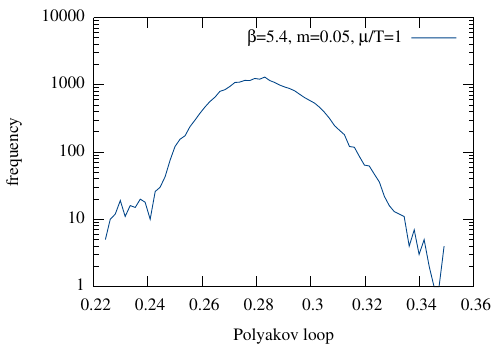}}
\centerline{\includegraphics{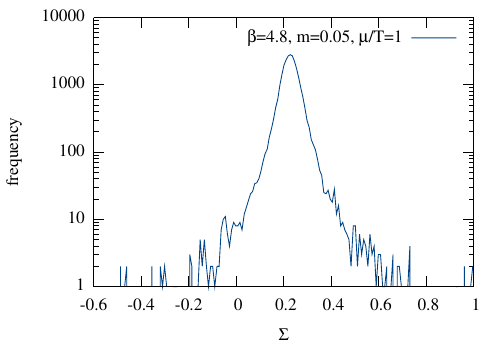}\includegraphics{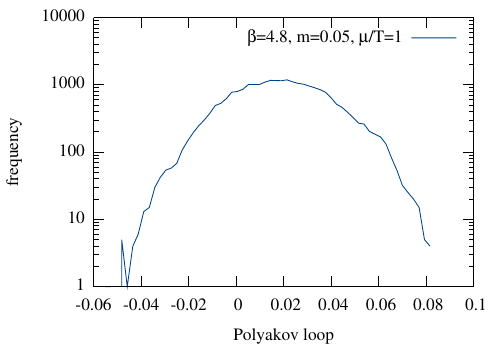}}
\caption{\label{fig:CLhist}Histogram of the chiral condensate (left) and Polyakov loop (right) for $\beta=5.4$ (top row) and $\beta=4.8$ (bottom row) for $m=0.05$ and $\mu=0.25$ on an $8^3\times 4$ lattice.\vspace{-3mm}}
\end{figure}

The fact that a stable solution is found does not necessarily mean that it is correct, as the CL method can converge to the wrong solution in some instances. In order to validate the CL results we show the histograms for the chiral condensate and the Polyakov loop for $\beta=5.4$ and $\beta=4.8$, in Fig.~\ref{fig:CLhist}.
The CL results are only to be trusted if the tails of the histograms decay exponentially. For the chiral condensate it seems that the validity might be problematic for $\beta=4.8$, as the tails of the histogram are somewhat broad, but even for $\beta=5.4$ the exponential decay in the tails is not so clear, even though the results agree with the reweighting results. The histograms for the Polyakov loop look fine so far. From these data we conclude that the results may be incorrect below the phase transition, even though the histograms do not give a clear cut way to validate or invalidate CL measurements.

\subsection{CL evolution for varying $\mu$}

As we found stable CL evolutions when decreasing $\beta$ through the phase boundary, we also performed a partial study of the phase diagram and investigated QCD for varying $\mu$ at two values of the temperature. 
We chose $\beta=5.0$ and $m=0.05$ for which the pion mass is $a m_\pi=0.5588 \pm 0.0002$ and $a=\unit[(0.3045 \pm 0.0001)]{fm}$, such that $m_\pi \approx \unit[362]{MeV}$. We work on an $8^3\times N_t$ lattice and consider temporal extents $N_t=4$ and $N_t=8$, corresponding to $T=\unit[161.74]{MeV}$ and $T=\unit[80.87]{MeV}$, respectively.

\begin{figure}
\centering
\includegraphics{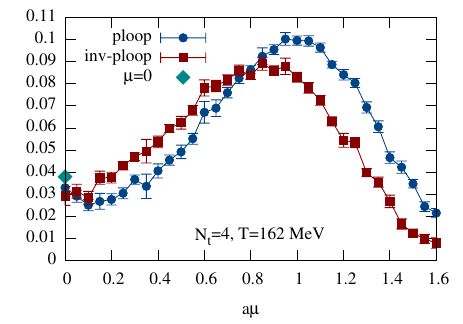}\includegraphics{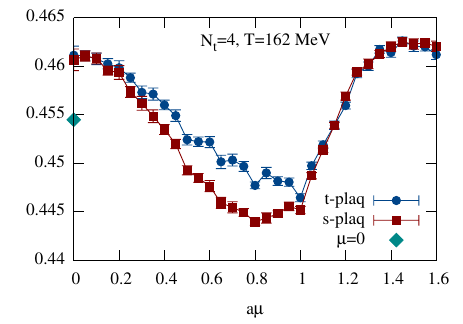}\includegraphics{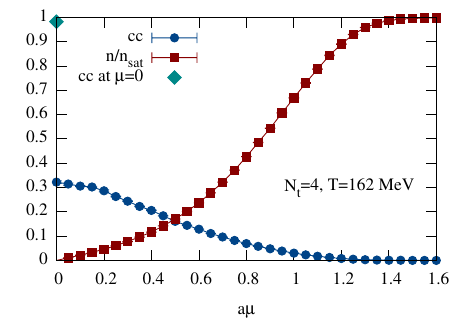}\\
\includegraphics{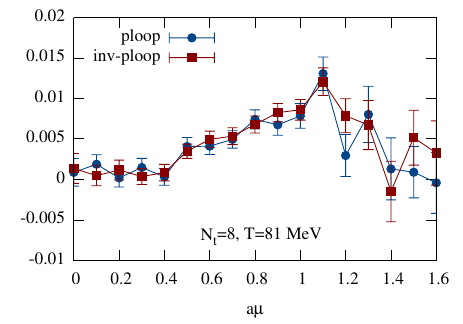}\includegraphics{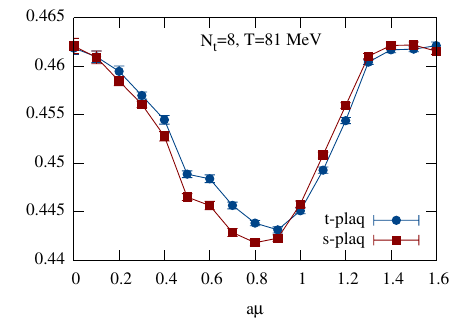}\includegraphics{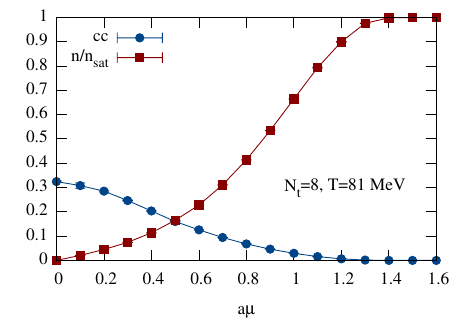}
\vspace{-2mm}
\caption{\label{fig:mu-evolution}Observables for $N_t=4$ (top) and $N_t=8$ (bottom):
Polyakov loop and inverse Polyakov loop (left), temporal and spatial plaquettes (middle), chiral condensate and quark number density (right) as a function of $\mu$ for $\beta=5.0$ and $m=0.05$. The exact $\mu=0$ results for $N_t=4$ were provided by Denes Sexty.\vspace{-6mm}}
\end{figure}

In Fig.~\ref{fig:mu-evolution} we show the results for the various observables as a function of $\mu$ for both temperatures. Although the results do not show any conspicuous behavior, there are no other results to compare with when the sign problem becomes large at nonzero $\mu$. One value that can easily be checked is the value at $\mu=0$ as it can be computed using standard importance sampling method. From the figure it is clear that the chiral condensate and even the plaquette measured from the CL simulations is incorrect for $\mu=0$, even though they result from a convergent CL evolution. For $\mu=0$ the Langevin evolution should be real, and the wrong result is merely due to numerical inaccuracies. This is in fact easily remedied by reunitarizing the links after each Langevin step; however, we decided against this to be consistent with the $\mu\neq0$ simulations. It is interesting to note that the measurements at nonzero chemical potential smoothly connect to the wrong $\mu=0$ value, and so we expect all of them to be incorrect. This argument is further supported through the lack of Silver Blaze phenomenon, as one would have expected a very slow $\mu$-evolution of the observables up to the phase transition.
The validity of the CL results can again be verified using the histograms of the measured observables. The histograms are very similar to those of the bottom row of Fig.~\ref{fig:CLhist}: that of the chiral condensate does not seem to have the required exponential falloff, while that of the Polyakov loop does not show a problem. Still, the histograms are not what one would brand as extremely broad, and so the decision of the validity is a difficult call, even though the chiral condensate at $\mu=0$ is wrong by a factor three.

\section{Summary and outlook}

In this presentation we have argued that the breakdown of the CL method at the QCD phase boundary observed in \cite{Fodor:2015doa} is in fact a numerical artifact due to the stochastic estimation of the drift. To compute the drift exactly, one needs the exact inverse of the Dirac operator, which cannot be computed in full with standard direct methods because it is too expensive in terms of both computer time and storage. However, we showed that the drift and the fermionic observables only require those elements of the inverse Dirac operator at the positions where the Dirac operator itself is filled, and, therefore, the selected inversion method, implemented in the sparse direct solver library PARDISO, can be applied. This allows for the exact computation of the drift term in a much faster way, using little storage space. 

We observed that the Langevin evolution became stable and convergent when the exact drift term was used. This allowed us to study the QCD phase transition across the roof of the phase diagram, i.e., when decreasing temperature from the deconfined to the confined phase at constant $\mu/T=1$.
Moreover, we were able to measure QCD observables as a function of $\mu$ at two values of the temperature below the deconfinement temperature.

Although these are preliminary results, there are clear indications that the CL results obtained at small mass in the confined region are incorrect, even though they are stable. There is therefore a need for further investigation of these CL results to understand if the validity problems are of a fundamental or numerical nature. The fact that we now get stable CL trajectories should allow us to dig deeper into this problem and to search for improved methods.

\subsection*{Acknowledgments}
{\small
This work was supported by the DFG collaborative research center SFB/TRR-55. We would like to thank Falk Bruckmann and Piotr Korcyl for useful discussions, Denes Sexty for providing the zero $\mu$ data, Radim Janalík for helping with the computer resources, and the ICS, Lugano for providing computing time for the simulations.}

\bibliography{biblio.bib}

\end{document}